\documentclass[prl,onecolumn,floatfix,superscriptaddress]{revtex4}
\usepackage{graphicx}
\usepackage{amsmath}

\graphicspath{ {./figurespdf/} } 

\begin{document}

\title{Understanding the structure of the first atomic  contact in Gold}

\preprint{1}

\author{C. Sabater}
\email{carlos.sabater@ua.es}
\affiliation
{Departamento  de  F\'{\i}sica Aplicada,
Universidad de Alicante, San Vicente del Raspeig, E-03690 Alicante, Spain}

\author{M.J. Caturla}
\affiliation
{Departamento  de  F\'{\i}sica Aplicada,
Universidad de Alicante, San Vicente del Raspeig, E-03690 Alicante, Spain}

\author{J. J. Palacios}
\affiliation
{Departamento de F\'{\i}sica de la Materia Condensada, 
Universidad Aut\'{o}noma de Madrid, Madrid E-28049, Spain}

\author{C. Untiedt}
\affiliation
{Departamento  de  F\'{\i}sica Aplicada,
Universidad de Alicante, San Vicente del Raspeig, E-03690 Alicante, Spain}


\begin{abstract}

 We have studied experimentally the phenomena of jump-to-contact (JC) and 
jump-out-of-contact (JOC) in gold electrodes. 
JC can be observed at the first contact when the two metals approach each other while JOC occurs in the last contact before breaking. 
When the indentation depth between 
the electrodes is limited to a certain value of conductance, a highly reproducible
 behaviour in the evolution of the conductance can be obtained 
for hundreds of cycles of formation and rupture. Molecular dynamics simulations of this process show how the two metallic electrodes are shaped 
into tips of a well-defined crystallographic structure formed through a mechanical annealing mechanism.
 We report a detailed analysis of  the atomic configurations obtained 
before contact and rupture of these stable structures and obtained their conductance using first-principles
quantum transport calculations. These results 
help us understand the values of conductance obtained experimentally in the JC and JOC phenomena 
and improve our understanding of atomic-sized 
contacts and the evolution of their structural characteristics. 
 
\end{abstract}


\maketitle

\section*{Keywords}

Electronic Transport, atomic size contacts, Mechanical Annealing, Jump to contact, Jump out of the contat, Molecular Dynamics Simulations, Ab-Initio, DFT.

\section*{Background}

Metallic atomic sized contacts can be created by Scanning Tunneling Microscopy (STM) \cite{AgrLev03,Pascual} or by mechanically controlled break junctions (MCBJ) \cite{AgrLev03,Muller}. 
In such nanocontacts the electrical conductance  is closely related to their minimum cross section. Therefore, by recording the conductance  while the electrodes are displaced with respect to each other (traces of conductance),
 one can infer the atomic 
structure of these contacts. However, to understand the structures formed at the contact 
it is necessary to make use of theoretical models. Uzi Landman  et al. \cite{Uzi} pioneered the  
use of molecular dynamics (MD) simulations 
to follow the variation of the minimum cross section  during  the process of stretching of a nanocontact. 
Later, Untiedt et al. \cite{j2c}, by  experimentally studying
the phenomena of jump-to-contact (JC) in gold and combining MD and electronic transport calculations, 
were able to identify the formation of three 
basic structures before contact between the two electrodes, although a limited analysis  
on the conductance values was presented there.

Trouwborst et al. \cite{Trouwborst} have also studied the phenomena of JC and JOC using 
indentation loops where the maximum conductance
was limited to $1 G_0$, where $G_{0} = 2\frac{e^2}{h}$ (quantum of conductance). These experiments showed 
that the elasticity of
the two electrodes is one of the relevant parameters to explain these phenomena. Despite  the above, there is not yet a unique picture that correlates the experiments  with the MD and transport calculations, regarding the different atomic structures that can be found at the contact.

On the other hand, experiments, together with molecular dynamics and electronic transport calculations based on density functional theory, show 
how very stable structures can be obtained by repeated indentation. This has been 
described as a mechanical annealing phenomenon  \cite{SabaterMec}. 
Limiting the maximum conductance value ($5G_0$ for gold) in the process of formation and rupture of a 
nanocontact leads to reproducible and atomically sharp pyramidal electrodes. 
This technique has recently been used by other authors \cite{Castellanos} to prepare tips in situ for 
low temperature STM.

In this paper we show experimental results of the JC and JOC phenomena for gold, analyzed simultaneously. 
We study the most probable configurations before the formation and breaking of nanocontacts with pyramidal form 
obtained from MD simulations emulating the process of mechanical annealing. As found earlier \cite{j2c}, the  contacts can be  classified into monomer, dimer and double contact. 
In order to correlate with the experimentally obtained conductance values, 
we calculated the conductance of these structures using first-principles
quantum transport models.

\section*{Methods}

 We have used a STM, where the tip and sample were two gold electrodes with purity $99.999\% $. 
The experiments were done at $4.2$K and cryogenic vacuum atmosphere. In order to obtain the conductance of the contacts the electrical current was measured while applying 100 mV constant bias voltage between the gold structures. Figure 1A shows traces of conductance in a gold nanocontact, measured in units of $G_0$ during the process of formation (red) and rupture (green). Insets show some snapshots 
from  our molecular dynamics simulations. These correspond to the initial structure (top figure) and 
the final structures before breaking (bottom right) 
and  just after contact formation (bottom left). Figure 1B is a zoomed area around $1 G_0$ of figure 1A, 
where the phenomena of JC and JOC can be clearly observed. 
In order to quantify the jump occurring in these two processes we define two 
conductance values for JC ($G_a$, $G_b$) and two values for JOC ($G_c$ $G_d$). 
These values correspond to the conductance values before and after the jump. We have performed thousands of indentations and recorded the values of these points. 
Representing  $G_b$ vs $G_a$  for the JC case and $G_d$ vs $G_c$ for JOC, we can obtain a colour 
density plot as shown in Figure 1C for JC and in Figure 1D for JOC. 
Lighter colours are less probable values than darker colours.
\begin{figure}[h]
\begin{center}
\includegraphics[scale=0.75]{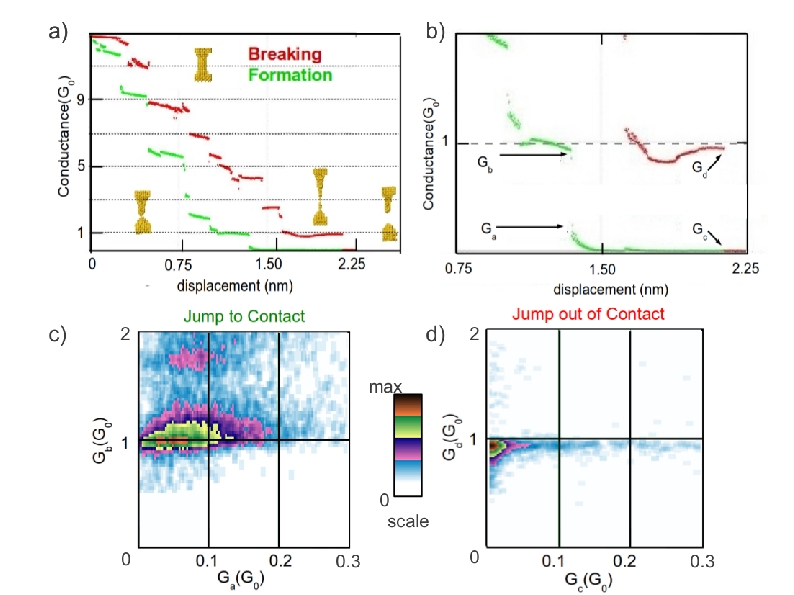}
\caption{ How to build a density plot.
    On the top left hand side we show a typical trace of conductance of gold at $4K$ during the process of breaking (red) and forming (green) a contact. The top right hand size is a zoom near $1 G_0$
to define the values before and after the jump-to-contact, $G_a$  and  $G_b$, and jump-out-of-contact, $G_c$  and  $G_d$. The bottom figures show colour density plots where dark colours represent those values of conductance that appear more frequently (left for JC and right for JOC).}
\end{center}
\end{figure}
 To emulate the movement of the STM 
and simulate the tip and surface annealed mechanically, we used MD simulations with embedded atom potentials. Density function theory (DFT) based calculations are performed to obtain 
the electronic transport in the simulated structures \cite{Alacant}. For the MD simulations we have selected an embedded atom potential \cite{Zhow2001} 
because elasticity of the electrodes seems to be one of the key parameters in the processes to be studied \cite{Trouwborst}, and these empirical potentials 
are fitted to reproduce the experimental elastic properties of bulk materials.

Furthermore the computational cost with this simulation method is low, which makes it 
an appealing tool since we need to simulate tens of these cycles of breaking and formation of the nanocontact. 

Using MD we have analyzed the same structures described in detail in reference \cite{SabaterMec}, but now we focused on the type of contact formed. 
The two initial configurations of the  nanocontacts  are shown in Figure 2. Structure A is built with 525 gold atoms.
This initial structure is stretched until the contact is broken by displacing the two top and bottom atom layers (represented in blue in the figure). 
After breaking, the direction of the displacement of these layers is reversed so that the two sides are brought together until contact. The temperature
in the simulations is $4.2K$. In this case the temperature is scaled in every cycle of breaking and formation of the contact. The indentation process continues
 until the minimum cross-section formed has 15 atoms. Then, the whole cycle starts again, breaking and forming the contact for a total of 20 cycles 
(see movie1 of supplementary material on ref. \cite{SabaterMec}).
The second structure studied (B) is shown in Figure 2, composed of 2804 gold atoms. 
In this case the indentation is limited to cross-sections of 
25 and 15 atoms (movie2 and movie3 at supplementary material on ref. \cite{SabaterMec}). The temperature here is kept constant and equal to $4.2K$ during 
the whole simulation by scaling the velocities of all atoms every time step (every femtosecond). The strain rates applied 
are between $10^{8}$ and $10^{10} s^{-1}$, typical of MD simulations \cite{Sorensen}. Note that the ratio of length of the contact to minimum cross-section 
is very different in these two structures (5 for structure A and 2 for structure B), therefore 
exploring a system with a long and narrow constriction and another of a short and wide nanowire. As shown previously \cite{SabaterMec}, structure A
reaches a stable configuration formed by two pyramidal tips after repeated indentations. This configuration is formed after cycle 11 and it remains stable 
for the following 9 cycles. In each of these cycles, although the pyramidal shape remains, there are differences in the atomic configurations right at the contact, 
as shown in Figure 3. These are the configurations we study and describe in this paper in detail. For the case of structure B, because of the initial shape, 
the formation of the two pyramidal tips occurs from the very first cycle and, again, only differences are observed in the very last atomic configuration forming the contact. 
We have performed electronic transport calculations based on DFT \cite{Gausian,Alacant} for both structures, A
and B.  These calculations have been carried out with the help of our code ANT.G, which is part of the package ALACANT \cite{Alacant} and implements the non-equilibrium Green's function formalism as a module to the popular code GAUSSIAN09 \cite{Gausian}. Here, due to the large number of atoms, we have employed a very basic basis set consisting only of one 6s orbital and one electron the remaining 78 electrons being part of the pseudopotential.

\begin{figure}[h]
\begin{center}
\includegraphics[scale=0.25]{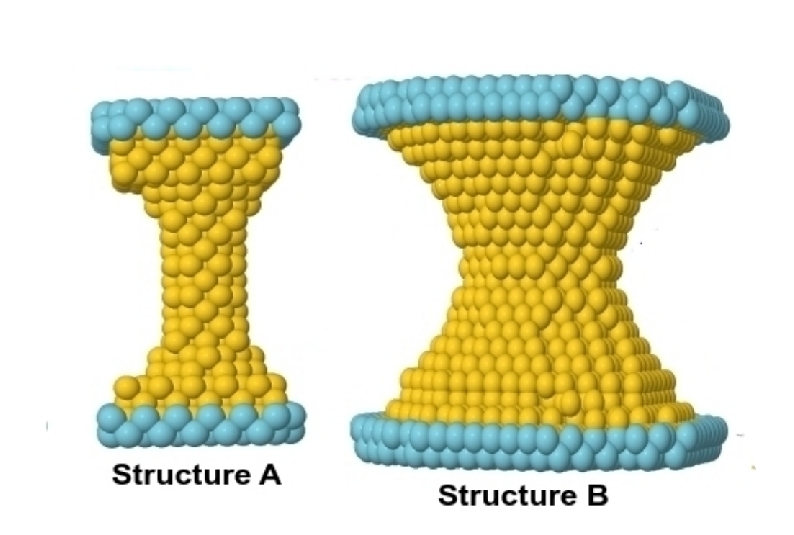}
\caption{Structure  and configurations of contacts.
The two initial configurations used in the MD simulations are shown: structure A, long and narrow contact and structure B, short and wide contact.}
\end{center}
\end{figure}

\section*{Results and discussion}

Experimental results of the JC and JOC in gold are shown in Figure 1 and Table I. Figures 1C and D show the color density plots obtained for gold when representing  $G_b$ vs $G_a$ for the case of JC, and $G_d$ and $G_c$ for the case of JOC. 
Note the presence of two very distinct areas in the JC plot, corresponding to configurations with a high probability.
In the case of JOC we can distinguish clearly one area of high probability. More details about these experiments are presented in reference  \cite{j2c}.
For clarity, we include in Table I those pairs of conductance that appear more frequently in the experimental measurements.
We should mention that for all traces studied in gold the phenomena of JC or JOC are always observed, 
unlike in other metals\cite{j2c}. 
For JC we observe three pairs of values that occur with higher frequency which we named as maximum 1, 2 and 3. 
In the JC maxima 1 and 2 correspond to jumping from a value of $0.01 G_{0}$ to a value of $0.94 G_{0}$ 
and from a value of $0.05 G_0$ to $0.98G_0$. These two peaks
are easily observed in Figure 1C as one large area of high probability. The last maximum corresponds to a jump from $0.09 G_0$ to $1.77 G_0$, which is the second spot
shown in Figure 1C. On the other hand, on breaking the nanocontact, only two maxima have been identified, 
one where the contact breaks
for conductance values of $0.92 G_0$, clearly seen in Figure 1D, and another one when it breaks 
at conductance values of $1.60 G_0$, which appears very faint in the figure. 
Note that these two values are close to those obtained for the first and third maximum in the JC case.

\begin{table}[h]
\caption{Experimental Values of conductance that appear more frequently in the case JC and JOC.
Pairs of values ($G_a$,$G_b$) and
($G_c$,$G_d$) for JC and JC-JOC respectively, that appear more frequently in the density plot of Figure 1.}
\centering
 \par \mbox{}
   \par
   \mbox{
     \begin{tabular}{|c|c|c|c|}
       \hline \multicolumn{4}{|c|}{Pairs of values obtained  in the density plots in fig.1}\\ \hline
       Phenomena &  1 maximum  &  2 maximum & 3 maximum \\ 
         & $ (G_a,G_b) G_0$  & $(G_a,G_b) G_0$ &$ (G_a,G_b) G_0$ \\ \hline
	    JC & (0.03,0.94)  & (0.05,0.98) &(0.09,1.77) \\ \hline
	    JOC & (0.01,0.92) & -&(0.01, 1.60)   \\ \hline
            \end{tabular}
     }
     \label{table1}
\end{table}

As mentioned above, we make use of molecular dynamics simulations and DFT calculations of 
conductance to understand these experimental measurements and observations.
Figure 2 shows the two structures studied using MD, as described earlier. In Figure 3 we show some 
snapshots of the configurations
found just after the contact between the two tips and just before breaking a nanocontact. 
Three basic atomic structures are found: a monomer and a dimer (Figures 3a and 3b),
and a double contact (Figures 3c-3e). For the case of a double contact we have identified 
different geometries, three of which are shown in this figure.
We introduce, for first time, the concept of a double dimeric (Figures 3c and 3d) and monomeric (Figure 3e) 
contact. We define a double dimeric contact 
as the one where the contact is between two atoms facing two other atoms, 
while we define a doble  monomeric contact as a contact where two atoms are contacting each other. 
Another interesting point is that for the double dimeric contact we have identified two possible structures, 
one where two atoms are perpendicular to the other two (Figure 3c), which we call transversal configuration (D.C. Dimeric T), and one where two atoms are parallel to the other two (Figure 3d), which we 
call parallel configuration (D.C. Dimeric P).

\begin{figure}[h]
\begin{center}
\includegraphics[scale=0.5]{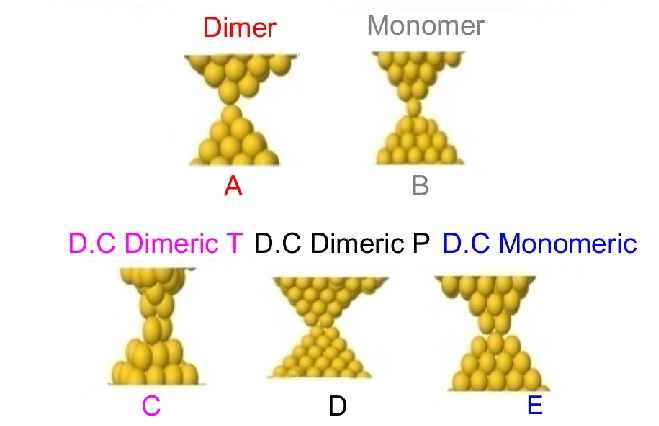}
\caption{Structures are the point of contact or before breaking from MD simulations.Representative configurations obtained from MD simulations right before contact or right before breaking are shown: (a) dimer (b) monomer 
(c) double contact dimeric transversal (d) double contact dimeric parallel and (e) double contact monomeric.}
\end{center}
\end{figure}

Table II shows the probability of finding a monomer, a dimer, or a double contact 
(all possible configurations for D.C) in the MD simulations right before contact and right after
contact, for the 2 initial structures and different indentations. Note the limited statistics in these results, 
since only 10 cycles have been computed for the first structure
and 9 cycles for the second one. Nevertheless, we can see some interesting results. 

\begin{table}[h]
\caption{MD results of the type of first contact or last contact (JC or JOC) in the structures A and B annealed mechanically. This table contains the probabilities of finding a dimer, monomer or double contact in the MD simulations for the two structures studied A and B during the formation or the breaking of the contact and for different indentation values (15 atoms or 25 atoms in the minimum cross-section).}
\centering
\par \mbox{}
   \par
   \mbox{
     \begin{tabular}{|c|c|c|c|}
       \hline \multicolumn{4}{|c|}{Percentage of cases of type Monomer, Dimer and D.C.}\\ \hline
       Phenomena & Monomer & Dimer  & Double Contact \\ \hline
	    JC  Structure A 15 iden & 20 & 60 & 20 \\ \hline
	    JOC Structure A 15 iden  &30 &60 &10 \\ \hline
	    JC  Structure B 15 iden & 0 & 0 & 100 \\ \hline
	    JOC Structure B 15 iden & 0 & 0 & 100 \\ \hline
	    JC  Structure B 25 iden & 22 & 0 & 78\\ \hline
	    JOC Structure B 25 iden & 22 & 56 & 22 \\ \hline
     \end{tabular}
     }
 \label{table2}
\end{table}

For the case of structure A, 
with a large ratio of length to minimum cross-section, we observe
that the most probable configuration both at JC and at JOC is a dimer. The monomer and the double contact 
have similar probabilities. This result is in agreement with reference
\cite{Wang}. The situation for the second structure, B, with a small ratio of length to minimum cross-section, 
is significantly different. In this case, when the indentation between the two 
tips is limited to 15 atoms in cross-section, the configuration at the contact is the same in all cycles, 
a double contact, although we observe the formation of the different
double contacts described in Figure 3c-e. Clearly, very stable pyramidal structures are formed in this case. The 
robustness of the tip imposes the repetition of a certain kind of structures. When the indentation
between the two tips increases to a value of 25 atoms in cross-section we should note that the traces 
do not repeat between cycles, and therefore, different structures are formed.
In this case, for JC, the double contact is still predominant, while for the JOC the probabilities have 
the same trend as in structure A (dimer being the most probable).

In order to correlate the results from molecular dynamics to the experimental measurements it is necessary 
to calculate the conductance of these atomic structures. 

\begin{table}[h]
\caption{Electronic conductance calculated by DFT on typical contacts obtained from MD structures.}
\centering 
 \par \mbox{}
   \par
   \mbox{
     \begin{tabular}{|c|c|c|c|}
       \hline \multicolumn{4}{|c|}{Structure and Value of Conductance $G_0$}\\ \hline
       Metal &Dimer& Monomer  & Doble Contact \\ \hline
	    Au&   0.92 $\pm$ 0.07 & 0.97 $\pm$ 0.15  & 1.73 $\pm$ 0.02 \\ \hline
           \end{tabular}
    }
 \label{table3}
\end{table}

Table III shows the values of condctance obtained from electronic transport calculations based on DFT 
for the typical first or last contacts proposed: monomer, dimer 
and doble contacts. The table includes the values of conductance obtained with their standard deviation.
We can observe that the monomer values of conductance are in the range  $1.20-0.76 G_0$, 
with an average value of $0.97 G_0$. 
That is because during the process of rupture and formation the monomer 
can be localized closer or further away from the rest of the contact. Another important factor that
can change the conductance of a monomer is the total number of neighboring atoms to the central atom in the contact, which can be different while remaining a monomer structure. 
Both factors are responsible for the spread in the conductance values of a monomer. 
On the other hand, the deviations in the conductance values  for dimer or double contact 
structures are significantly smaller, around $0.07G_0$ and $0.02G_0$ respectively, 
the average conductance value being
$0.92G_0$ for a dimer and $1.73G_0$ for a double contact. These results indicate that, on average, dimer and monomers have similar values of conductance while double contacts have
significantly larger conductance values. It seems clear then that the maximum obtained 
experimentally for JC and JOC, with conductance values of $1.77G_0$ and $1.6G_0$ respectively,
(maximum 3 for JC and maximum 2 for JOC in Table I) correspond to the formation of a double contact. 
The results for the other maxima obtained experimentally are not so clear since the average
conductance values obtained for a monomer and a dimer in the calculations are very similar. 
This seems to indicate that the two first maxima obtained experimentally in the 
JC must correspond to configurations in a dimer and in a monomer geometry. 
According to MD simulations, the most likely configuration both in JC and 
JOC is a dimer (except in special cases of very stable tips), although monomers can also be formed.

Figure 4 combines the experimental results with the calculations of conductance in order 
to identify the experimental values of JC and JOC with structures detected by MD 
and calculated by DFT. The center column between the experimental density plots of JC and JOC indicates the average value of conductance obtained from the simulations 
for each geometry (double contact, monomer and dimer). The thickness of the rectangles 
around each geometry indicates the standard deviation. It is clear from this plot that
the top high frequency events in the density plots corresponds to a double contact and the bottom 
high frequency events corresponds to monomer and dimer configurations.
Although, as we mentioned above,  it is difficult to distinguish the monomer and dimer using our 
theoretical model, we can see that the average of conductance of monomers is above the one of the dimers. 
If we add to this that we would expect a higher tunnel conductance (on average) prior to the formation of a 
monomer we can label maximum 1 and 2 as dimer and monomer respectively.

\begin{figure}[ht]
\begin{center}
\includegraphics[scale=0.75]{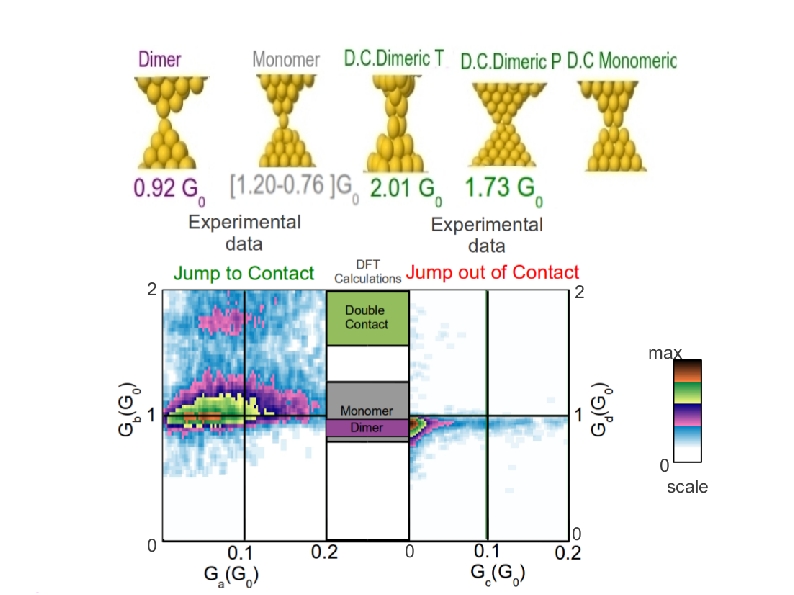}
\caption{JC and JOC density plots together with conductance calculations of different geometries of the contact.Inside the experimental density plots we have marked the average conductance values after or before the jump as obtained from DFT electronic transport
calculations with their deviations.}
\end{center}
\end{figure}

\section*{Conclusions}

Experiments of JC-JOC show that certain structures are more likely to occur than others. This depends on the metal and on the process of breaking/formation and 
the type of structure at the electrodes. 
Simulations and calculations (MD and DFT) of these experiments show that three basic atomic structures are formed at the contact: monomers, dimers and double contacts. 
We have identified within the double contact structure several different atomic arrangements that we named double dimeric contact (parallel and perpendicular), 
and doble monomeric contact. According to DFT electronic transport calculations, double contacts have 
an average value of conductance of $1.73G_0$, which correlate very well
with one of the peaks observed experimentally both for JC and for JOC. This configuration is also obtained in JC and JOC from the MD simulations and, for some very stable tips,
is the dominant configuration. Monomers and dimers, however, are difficult to distinguish from the 
simulations since their average conductance values are very similar 
($0.97G_0$ and $0.92G_0$ respectively). In the case of JOC these two peaks can not be resolved. Interestingly,
 the  conductance values are somehow lower than in the case of JC, 
which could indicate the most likely formation of stretched contacts.

\newpage

\section*{Author's contributions}
  CS wrote the manuscript, did MD simulations  and  DFT calculations. CU and CS performed the experiments. MJC and JJP supervised the MD and DFT calculations. All the authors have participated in the outline of this research, in the bibliographical study  and revised the manuscript. All authors read and approved the current manuscript.
\section*{Competing interests}
   The authors declare that they have no competing interests.

\section*{Acknowledgements}
Work supported by the Spanish government through grants FIS2010-21883, CONSOLIDER CSD2007-0010,  Generalitat Valenciana through PROMETEO/2012/011, ACOMP/2012/127 and Feder funds from E.U.





\begin{thebibliography}{10}
 
\providecommand{\url}[1]{[#1]}
\providecommand{\urlprefix}{}
\bibitem{AgrLev03}
N. Agrait, A. Levy-Yeyati, J. M. van Ruitenbeek, \textbf{Quantum properties of atomic-sized conductors} \emph{Phys. Rep}. 2003, \textbf{377} 81 

\bibitem{Pascual} 
 J.I. Pascual, J. Mendez, J. Gomez-Herrero, A.M. Baro, and N. Garcia,\textbf{Quantum Contact in Gold Nanostructures by Scanning-Tunneling-Microscopy} \emph{Phys. Rev. Lett} 1993 \textbf{71}, 1852

\bibitem{Muller} 
C.J. Muller, J.M. van Ruitenbeek, and L.J. de John, \textbf{Conductance and supercurrent discontinuities in atomic-scale metallic constrictions of variable width} \emph{Physica C}1992, \textbf{191}, 485

\bibitem{Uzi}
U. Landman, W. D. Luedtke, N. A. Burnham and R. J.Colton \textbf{Atomistic mechanisms and dynamics of adhesion, nanoindentation, and fracture} \emph{Science} 1990, \textbf{248} 454 




\bibitem{j2c}
C. Untiedt, M. J. Caturla, M. R. Calvo, J. J. Palacios, R. C. Segers  and J. M. van Ruitenbeek,\textbf{Formation of a Metallic Contact: Jump to Contact Revisited} \emph{Phys. Rev. Lett.} 2007, \textbf{98} 206801

\bibitem{Trouwborst}
M. L. Trouwborst, E. H. Huisman, F. L. Bakker, S. J. van der Molen and B. J. van Wees \textbf{Single Atom Adhesion in Optimized Gold Nanojunctions} \emph{Phys. Rev. Lett.} 2008, \textbf{100} 175502

\bibitem{SabaterMec}
C. Sabater, C. Untiedt, J. J. Palacios, and M.J. Caturla, \textbf{Mechanical annealing of metallic electrodes at the atomic scale} \emph{Phys. Rev. Lett.} 2012, \textbf{108} 205502 


\bibitem{Castellanos}
Andres Castellanos-Gomez, Gabino Rubio-Bollinger, Manuela Garnica, Sara Barja, Amadeo L. Vazquez de Parga, Rodolfo Miranda, Nicolas Agraıt, \textbf{Highly reproducible low temperature scanning tunneling microscopy and spectroscopy with in situ prepared tips} \emph{Ultramicroscopy} 2012, \textbf{122} 1-5 

\bibitem{Alacant}
 ALicante Atomistic Computation Applied to NanoTransport. Package publicly available
at http://alacant.dfa.ua.es.

\bibitem{Zhow2001}
    X.W. Zhoua, H.N.G. Wadleya,R.A. Johnsona, D.J. Larsonb, N. Tabatb,A. Cerezoc, A.K. Petford-Longc,G.D.W. Smithc, P.H. Cliftond, R.L. Martense,
    T.F. Kellye \textbf{Atomic scale structure of sputtered metal multilayers} \emph{Phys. Rev. B} 2001,\textbf{49},4005–4015

\bibitem{Sorensen}
Mads R. Sørensen, Mads Brandbyge, and Karsten W. Jacobsen \textbf{Mechanical deformation of atomic-scale metallic contacts: Structure and mechanisms}
 \emph{Phys. Rev. B} 1998,\textbf{57},3283–3294



\bibitem{Gausian}
M. J. Frisch et al., Gaussian 09 revision a.1, Gaussian Inc. Wallingford CT 2009.

\bibitem{Wang}
Huachuan Wang,Yongsheng Leng \textbf{Molecular dynamics simulations of the stable structures of single atomic contactsin gold nanojunctions} \emph{Phys. Rev. B} 2011, \textbf{84} 245422

 \end{thebibliography}
\end{document}